\documentclass[aps,pre,preprint,showpacs,showkeys]{revtex4}
\usepackage{amssymb}
\usepackage[dvips]{color,graphicx}  
\usepackage{natbib}
\usepackage{color}

\begin{document}

\begin{flushleft}
 
{\Large {\bf Strategy for stopping failure cascades in interdependent
    networks}}
\newline
\\

Cristian E. La Rocca \textsuperscript{1,*},
H. Eugene Stanley \textsuperscript{2},
Lidia A. Braunstein \textsuperscript{1,2}
\\

\bigskip
{\bf 1} Departamento de F\'isica, Facultad de Ciencias Exactas y
Naturales, Universidad Nacional de Mar del Plata and Instituto de
Investigaciones F\'isicas de Mar del Plata (IFIMAR-CONICET), De\'an
Funes 3350, 7600 Mar del Plata, Argentina\\
\bigskip
{\bf 2} Physics Department and Center for Polymer Studies, Boston
University, Boston, Massachusetts 02215, USA\\
\bigskip

* Corresponding author\\
E-mail: larocca@mdp.edu.ar

\end{flushleft}

\begin{abstract}

Interdependencies are ubiquitous throughout the world. Every
real-world system interacts with and is dependent on other systems,
and this interdependency affects their performance. In particular,
interdependencies among networks make them vulnerable to failure
cascades, the effects of which are often catastrophic. Failure
propagation fragments network components, disconnects them, and may
cause complete systemic failure. We propose a strategy of avoiding or
at least mitigating the complete destruction of a system of
interdependent networks experiencing a failure cascade. Starting with
a fraction $1-p$ of failing nodes in one network, we reconnect with a
probability $\gamma$ every isolated component to a functional giant
component (GC), the largest connected cluster. We find that as
$\gamma$ increases the resilience of the system to cascading failure
also increases. We also find that our strategy is more effective when
it is applied in a network of low average degree. We solve the problem
theoretically using percolation theory, and we find that the solution
agrees with simulation results.
 
\end{abstract}



\maketitle

\section{Introduction}

Interdependence is a characteristic of the world in which we live. We
see this in such infrastructure networks as transportation systems,
electrical power grids, natural gas and water systems, telephone
systems, and the Internet. Thus, transportation networks depend on
petroleum supplies and electrical power, the Internet on electrical
power, and the control of natural gas and water systems on
telecommunications. Interdependencies among networks can produce new
systemic behaviors not seen in isolated networks, and these
interdependencies can both enhance network functioning or increase its
vulnerability to catastrophic failure. For example, a transportation
network can increase the propagation rate of a disease epidemic, and
if the network includes airlines the propagation can become
world-wide.

We see a catastrophic example of interdependence in the after-effects
of Hurricane Katrina in 2005
\cite{Chang_09,Rinaldi_01,Leavitt_06}. Several oil rigs and refineries
were destroyed and this paralyzed oil and gas extractions for a number
of months. This caused the price of fuel to rise exponentially, and
this affected the airlines. Forest devastation affected the
Mississippi logging industry, and there was a sharp drop in activity
in the ports of Southern Louisiana and New Orleans, two of the largest
in the EEUU. In addition to the thousands of homes that were
destroyed, the loss of thousands of jobs meant that many owners of
homes who had survived the hurricane were no longer able to pay their
mortgages. Some insurance companies, because of the huge indemnities
they had to face, either increased their homeowner fees or stopped
insuring in the area altogether. It is thus clear that we need to
understand how real-world interdependencies function. We need to know
how to prevent, avoid, or mitigate the catastrophic failures they can
magnify because interdependencies are everywhere.

Interdependent systems have recently been treated as networks of
networks (NoN), i.e., systems in which two or more networks interact,
and they have been successfully used to understand epidemic spreading
\cite{Granell_13_1,Alvarez-zuzek_15,Zhao_14,Boccaletti_14,Kiv_13},
failure cascades \cite{Boccaletti_14,Kiv_13,Val13,Reis_14,Scala_14},
diffusion \cite{Boccaletti_14,Kiv_13,Arenas_14,Gomez_13}, and
synchronization
\cite{Boccaletti_14,Hunt_08,Zang_15,Torres_15,Jin_11}. We characterize
single networks in terms of their internal degree distribution $P(k)$,
which is the probability that a node is connected to $k$, with $k_{\rm
  min} \le k \le k_{\rm max}$, where $k_{\rm min}$ and $k_{\rm max}$
are the minimum and maximum connectivities in the network,
respectively. The interdependence between networks involves
``external'' dependency links that connect nodes in one network to
nodes in a second network. These dependency links can strongly affect
system robustness and can facilitate such catastrophic events as
failure cascades
\cite{Boccaletti_14,Kiv_13,Val13,Reis_14,Scala_14,Bul_01} in which a
node failure in one network propagates through dependency links and
causes nodes in the other network to also fail. This occurred in the
power outage in Italy on 28 September 2003. The shutdown of some power
stations caused nodes in the communication network to fail, which in
turn caused breakdowns in additional power stations \cite{Rosato_08}.

In a model introduced by Buldyrev {\it et al.} \cite{Bul_01} the
authors studied the cascade failures in two interdependent networks by
mapping the process as random node percolation, a process that is very
important due to its ubiquitous application in failure cascades and
the spread of disease. In random node percolation a fraction $1-p$ of
nodes fail and the network fragments into clusters. The network
remains functional if there is still a connected giant component
(GC). The finite clusters remaining are considered dysfunctional. In
the Buldyrev model nodes in the first network depend one-to-one on
nodes in the second network. At the initial stage a fraction $1-p$ of
nodes and all finite clusters are removed from the first network and,
as a consequence, all the interdependent nodes in the second network
also fail. At each time step we remove all finite clusters and their
interdependent nodes in the other network until the system reaches the
state in which there are no remaining finite clusters. If both
networks still have functional clusters in this final ``steady''
state, they are of the same size and all their nodes are supported by
nodes in the opposite network. They found a threshold $p_c$ at which
all functional components disappear. This threshold is higher than in
isolated networks that have the same degree distribution, which
implies that a NoN is less robust than isolated networks. More
important is the nature of the phase transition characterized by the
order parameter $P_\infty(p)$ (the relative size of the GC), which is
first order while in isolated networks it is of second order
\cite{Sta_01}. In a first order transition the GCs overcome an abrupt
transition from a finite value to zero at $p_c$ and, as a consequence,
it is more difficult to forecast or control the transition than in
isolated networks.

After this pioneering work, many studies focused on modeling
mitigation strategies of preventing the drastic consequences of the
first order phase transition, such as autonomizing a fraction $q$ of
nodes
\cite{Pash_10,Pash2010,Sch_13,jia_02,Val13,Val14,Bul_11}. Nevertheless,
it is difficult and expensive to autonomize nodes in real NoN because
their infrastructure were constructed not at random, but instead by
economic reasons in order to accomplish efficiently some
tasks. Recently Di Muro {\it et al.} \cite{Dimuro_16} proposed a node
recovery strategy in a system of interdependent networks that repairs,
with probability $\gamma$, a fraction of failed nodes that are
neighbors of the largest connected component in each network. They
found that for a given initial failure of a fraction $1-p$ of nodes,
there is a threshold probability of recovery above which the cascade
stops and the system is restored to its initial state and below which
the system abruptly collapses. They found three distinct phases: one
in which the system never collapses without being restored, a second
in which the recovery strategy avoids collapse, and a third in which
the repairing process cannot prevent the collapse of the
system. However, it is not always possible to repair the components of
a network, and sometimes the repairing process requires so much time
and so many resources that the system suffers total breakdown before
it is completed. Thus, in this work we propose and study another
strategy for preventing total network destruction after a failure
cascade is initiated. The strategy is to save finite clusters prior to
their failure by connecting them to the functional network component
(GC).

\begin{figure}
\begin{center}
  \includegraphics[width=0.45\textwidth]{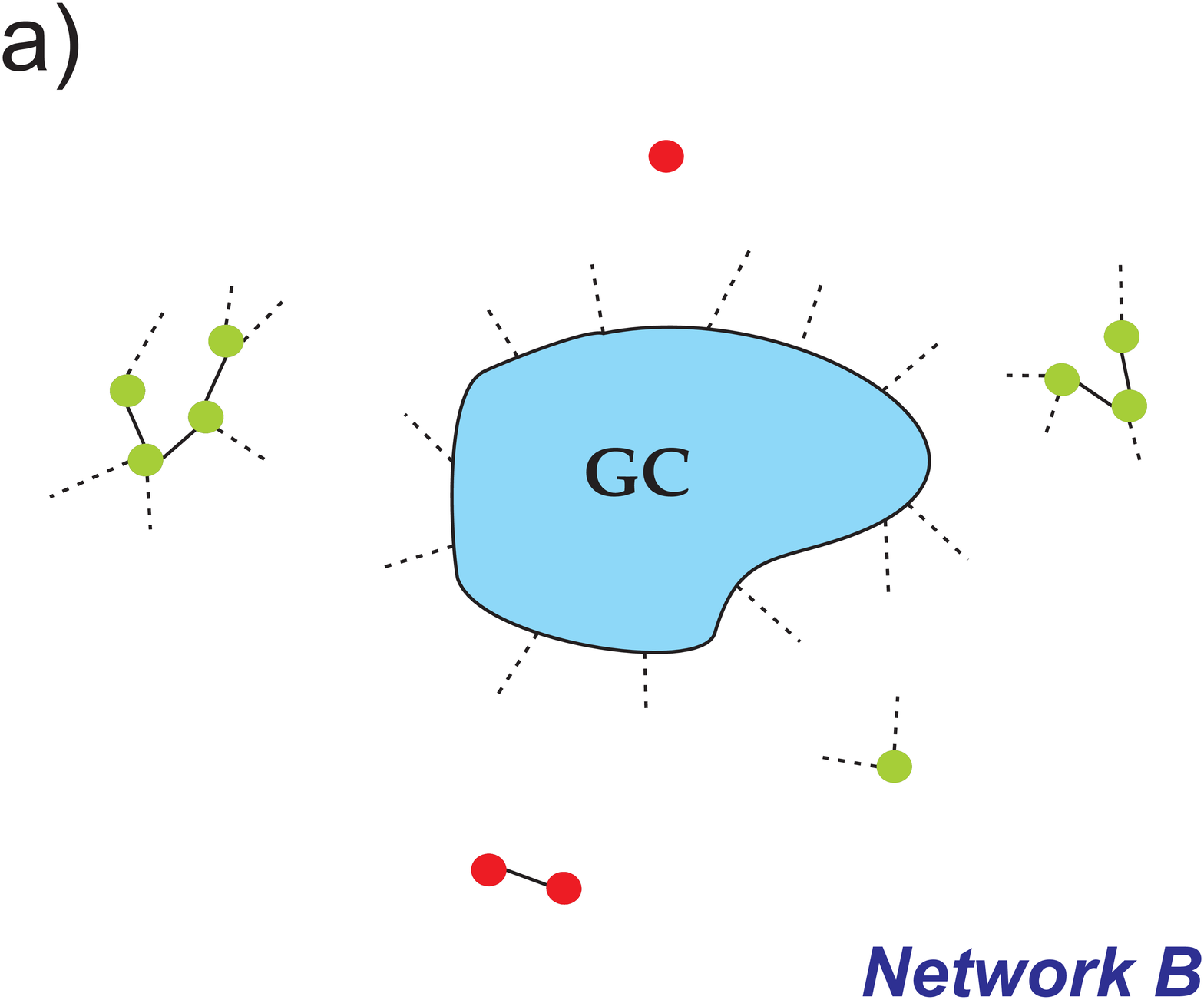}
  \hspace{1cm}
  \includegraphics[width=0.45\textwidth]{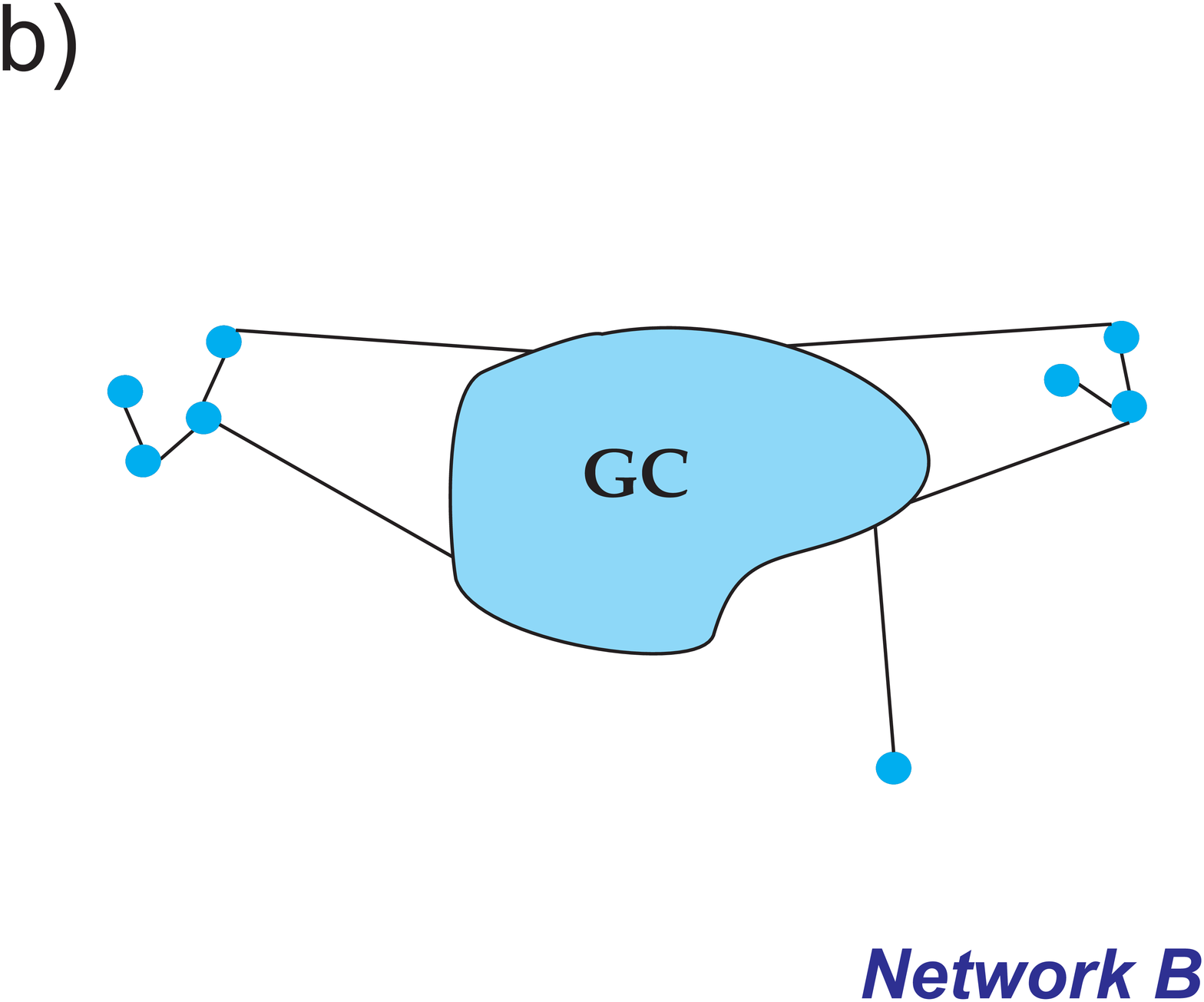}
\end{center}
\caption{Schematic of the implementation of the strategy in network
  B. The blue cluster represents the functional component (GC). $a)$
  the green clusters represent the saved clusters and in red the
  clusters that could not be saved. The dashed lines represent the
  free links available to use for the reconnection of the finite
  clusters. $b)$ The new GC of network B.}\label{fig_esq}
\end{figure}

\section{Model and simulations}

Using the process proposed by Buldyrev {\it et al.}  \cite{Bul_01} we
model and use a strategy for mitigating a cascade as it begins. We
consider two interdependent networks A and B with the same size $N$
and with degree distributions $P_A(k)$ and $P_B(k)$, respectively. We
use the Molloy-Reed algorithm \cite{Mol_01}, disallowing self loops
and multiple connections, to construct each network. The
interdependency between A and B we assume to be one-to-one, i.e., each
node has only one dependency link. We apply the strategy only to
network B---although it could be applied to either network---because
the cascading is so rapid that we must choose a network in which
applying the strategy is easier, such as a communication network,
which is faster and less expensive to operate than a power grid.

At step $n=0$ we remove a fraction $1-p$ of nodes in network A and
locate its GC. We then remove all nodes in A that belong to finite
clusters, assuming that they have become dysfunctional due to lack of
support, and we remove their interdependent nodes in B. We then apply
the strategy of locating the GC in B and reconnecting to it every
finite cluster in the same network with a probability $\gamma$ (See
Fig.~\ref{fig_esq}). We do this by connecting two nodes in each
existing finite cluster (clusters bigger than one node) with the
GC. Single nodes we connect using one connection. We select two nodes
rather than one when reconnecting to reduce the probability that the
finite cluster will again disconnect from the GC. Note that although
increasing the number of connected nodes increases network resilience,
a greater number of connections requires greater economic resources,
and cascading can be so rapid we are unable to achieve many
reconnections. Note also that in order to preserve the initial degree
distribution, the nodes we choose to reconnect must have free links,
i.e., links that at earlier stages were connected to nodes that were
already removed (dashed links in Fig.~\ref{fig_esq}). This assumption
allows us to map our model directly using node percolation theory. We
assume that the clusters that are finite prior to the failure cascade
will fail because they have no saveable free links.

The initial stage $n=0$ ends when we remove with a probability
$1-\gamma$ the finite clusters in B that cannot be saved [red nodes in
  Fig.~\ref{fig_esq}(a)]. Stage $n=1$ begins when we propagate the
failure from B back to A and remove all failed dependent nodes in B.
We iterate this procedure until the system reaches the final
``steady'' state in which there are no remaining finite clusters. We
denote by $P^\alpha_\infty$ the relative size of the GC in network
$\alpha$, with $\alpha=~$A,B. Note that because nodes in one network
depend one-to-one on nodes in the other network at the steady state,
$P^{\rm A}_\infty=P^{\rm B}_\infty=P_\infty$.  For the simulations we
use an Erd\"os R\'eny (ER) random graph characterized by a Poisson
degree distribution given by $P(k)=e^{- \langle k \rangle} \langle k
\rangle^k / k!$, where $\langle k \rangle$ represents the average
degree of the network, and a scale-free (SF) with cutoff with degree
distribution $P(k) \sim k^{-\lambda}exp(-k/\beta)$, where $\lambda$ is
the broadness of the distribution and $\beta$ is the cutoff in the
connectivity. For all simulations all networks have $N=10^6$ nodes,
with a maximum connectivity $k_{\rm max}=20$ and $\langle k \rangle
=8$ for the ER and $\lambda=2.5$, $k_{\rm min}=2$ and $k_{\rm
  max}=N^{1/2}$ \cite{Maslov_02,Xul_01,Xul_05} with $\beta=20$ for the
SF.
\begin{figure}
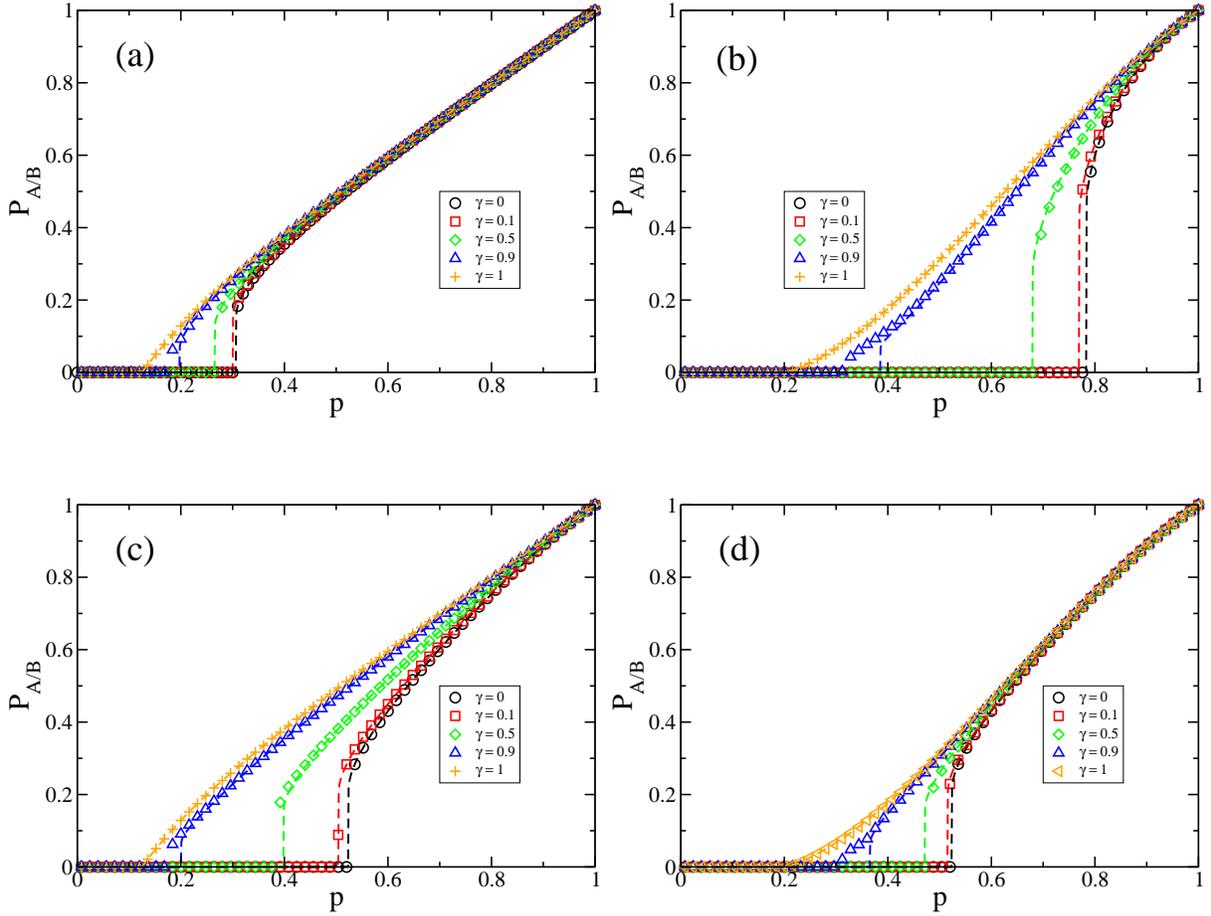

\vspace{1cm}
\begin{center}
\includegraphics[width=0.48\textwidth]{fig2a.eps}  
\includegraphics[width=0.48\textwidth]{fig2b.eps}\\
\vspace{1cm}
\includegraphics[width=0.48\textwidth]{fig2c.eps}
\includegraphics[width=0.48\textwidth]{fig2d.eps}
\end{center}
\caption{$P_\infty$ as a function of $p$ for $\gamma=0$ ($\bigcirc$),
  $\gamma=0.1$ ($\Box$), $\gamma=0.5$ ($\diamond$), $\gamma=0.9$
  ($\triangle$) and $\gamma=1$ ($+$) for $a)$ two ER, with $\langle k
  \rangle=8$ $b)$ two SF with $\lambda=2.5$ and $\beta=20$, and the
  combination of the two kind of network ER and SF applying the
  strategy in $c)$ SF and $d)$ ER. The dashed curves represent the
  theory.}\label{figura_1}
\end{figure}

Figure \ref{figura_1} plots $P_\infty$ at the steady state as a
function of $p$ for different values of $\gamma$. Note that for
$\gamma=0$ we recover the results of Ref.~\cite{Bul_01}, i.e., we
obtain a first-order transition in which $P_\infty$ jumps from a
finite value to zero at $p_c$. As $\gamma$ increases the threshold
$p_c$ decreases. This allows to the system to overcome a high level of
damage before the two functional components collapse. There is still a
first-order transition, but the jump in the relative size of the GC
decreases as $\gamma$ increases. Only when $\gamma=1$ is there a
continuous second-order transition. Note that the effect of the
strategy is stronger in networks with a low average connectivity, such
as for the SF networks which have an average degree $\langle
k\rangle=3.08$. Networks with a low average degree (the two SF) are
more likely to fragment, which makes the system more vulnerable to
failure cascades. Here and when $\gamma=0$ the system is completely
destroyed when 20 percent of its nodes fail, but using the strategy
the system can sustain a higher level of damage before it
collapses. The best value is $\gamma=1$. Here the system fails only
when 80 percent of its nodes fail. 

When the two networks are different our strategy continues to increase
systemic resilience to failure cascades, since the value of $p_c$
decreases as $\gamma$ increases. Note also that the best outcome
occurs when the strategy is applied to the more fragile network---the
one with the lowest average degree---in our case the SF network. Note
also that when $\gamma=0.5$ the ER-SF case with the strategy applied
to the SF network has a $p_c$ smaller than the SF-ER case with the
strategy applied to the ER network. When we do not use the strategy
($\gamma=0$), the $p_c$ is the same for both cases and is unaffected
by which network initiates the failure. However as $\gamma$ increases
the $p_c$ for ER-SF and SF-ER cases increasingly differ, because it is
more efficient to apply the strategy to the more fragile network,
which fragments more easily and in which unsaved finite clusters
ultimately fail.

\section{Analytical results}

Theoretically the cascading failure problem can be solved using node
percolation \cite{Bul_01,Jia_01}. In isolated networks we compute the
relative size of the GC after removing a fraction $1-\mu$ of nodes by
solving the self-consistent equation for probability $f$.  Choosing an
edge of the network at random leads to a node connected to the GC,
where $f$ is
\begin{equation}\label{fa}
f=\mu (1-G_1(1-f))\ ,
\end{equation}
where $G_1(x)=\sum_{k=k_{\rm min}}^{k_{\rm max}} k P(k) /\langle k
\rangle x̣^{k-1}$ is the generating function of the excess degree
distribution \cite{New_10,Jia_01}, $\langle k \rangle$ is the average
degree of the network, and $\mu$ is the effective fraction of
remaining nodes. Thus the fraction of nodes in the GC is
\begin{equation}\label{perc}
P_\infty(\mu)=\mu (1-G_0(1-f))\ ,
\end{equation}
where $G_0(x)=\sum_{k=k_{\rm min}}^{k_{\rm max}} P(k) x̣^k$ is the
generating function of the degree distribution
\cite{New_10,Jia_01}. For two interdependent networks, we start the
cascading at stage $n=0$ by removing a fraction $1-p$ of nodes in
network A ($\mu_A(0)=p$) and a corresponding fraction
$S_A(0)=\mu_A(0)-P_\infty^A(0)$ of nodes in the finite clusters. The
failure spreads to network B through interdependency links, thus
$\mu_B(0)=P_\infty^A(0)$ and the fraction of nodes in the finite
clusters in B is given by $S_B(0)=\mu_B(0)-P_\infty^B(0)$. The next
stage begins when the failure $S_B(0)$ of nodes in network B returns
to network A. For any step $n$ in the cascading, the stage begins when
a fraction $S_B(n-1)$ of nodes fail in the GC of network A
($P_\infty^A(n-1)$), which corresponds to the fraction of nodes
belonging to finite clusters in B at step $n-1$. Then the total
fraction of nodes that fails in network A at step $n$ is
$\mu_A(n-1)S_B(n-1)/P_\infty^A(n-1)$ and $\mu_A(n)$ is
\begin{equation}\label{muna}
\mu_A(n)=\mu_A(n-1)\left[1-\frac{S_B(n-1)}{P_\infty^A(n-1)}\right]\ .
\end{equation}
Thus we obtain $P_\infty^A(n)$ using Eqs.~(\ref{fa}) and (\ref{perc}),
and the fraction of nodes that belongs to the finite clusters as a
consequence of the fragmentation of $P_\infty^A(n-1)$ is
\begin{eqnarray}\label{sa}
S_A(n)=P_\infty^{B}(n-1)-P_\infty^A(n)\ .
\end{eqnarray}
Following the same procedure as in network A, we obtain
\begin{eqnarray}\label{munb}
\mu_B(n)=\mu_B(n-1)\left[1-\frac{S_A(n)}{P_\infty^{B}(n-1)}\right]\ ,
\end{eqnarray}
\begin{eqnarray}\label{sb}
S_B(n)=P_\infty^A(n)-P_\infty^B(n)\ ,
\end{eqnarray}
where $P_\infty^B(n)$ is derived using Eqs.~(\ref{fa}) and
(\ref{perc}). The process continues until the system reaches the steady
state in which there are no finite clusters in either network
$S_A=S_B=0$ and both GCs are of the same size $P_\infty^A=P_\infty^B$.
\begin{figure}
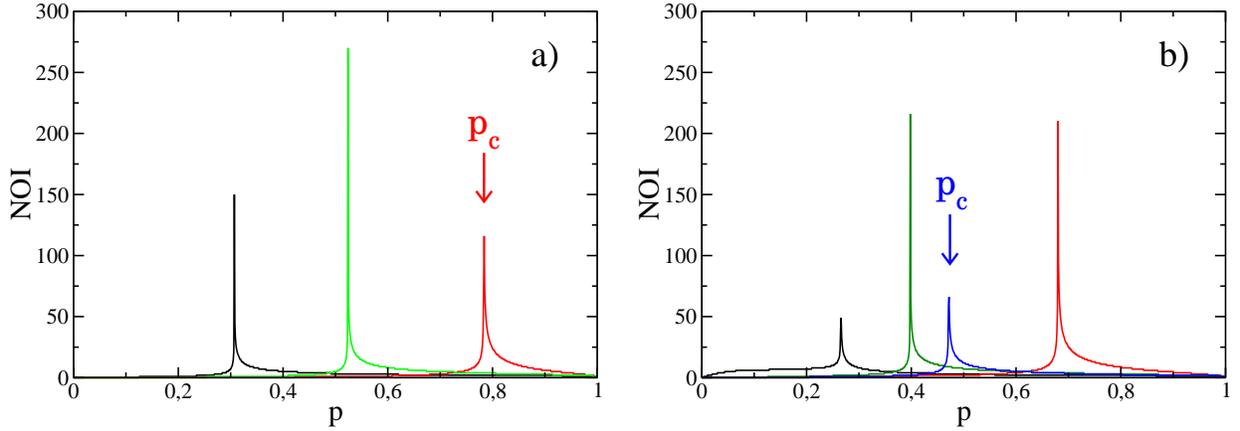
 
\vspace{1cm}
\begin{center}
  \includegraphics[width=0.48\textwidth]{fig3a.eps}
  \hspace{0.2cm}
  \includegraphics[width=0.48\textwidth]{fig3b.eps}  
\end{center}
\caption{NOI as a function of $p$ from the theory for a) $\gamma=0$
  and b) $\gamma=0.5$. Both cases are for two ER (black), two SF
  (red), ER-SF (green) and SF-ER (blue), always applying the strategy
  in B ($A-B$). Notice that in $a)$ the green curve represents the two
  cases ER-SF and SF-ER because without strategy the cascading failure
  is the same.}\label{figura_3}
\end{figure}

The strategy is applied after the failure in A propagates to B and
just prior to its return to A. We now save with a probability $\gamma$
each finite cluster in B. Even when finite clusters have differing
sizes and probabilities of existing, we assume that on average we save
a fraction $\gamma S_B$ of nodes at each time step because the
probability $\gamma$ of saving each finite cluster is independent of
its size (its number of nodes). We begin at stage $n$ in network A
with failure $(1-\gamma) S_B(n-1)$, which corresponds to the finite
clusters in B that could not be saved during the previous step. Thus
we rewrite Eq.~(\ref{muna}) to be
\begin{equation}
\mu_A(n)=\mu_A(n-1)
\left[1-\frac{(1-\gamma)S_B(n-1)}{P_\infty^A(n-1)}\right]\ . 
\end{equation}
The cascading process evolves following Eqs. (\ref{sa}), (\ref{munb}),
and (\ref{sb}), with Eqs. (\ref{fa}) and (\ref{perc}). We apply our
strategy and save a fraction of nodes $\gamma S_B(n)$. Then the
fraction of nodes belonging to the new GC in B is
\begin{eqnarray}\label{eqpb}
P_\infty^{B}(n)=P_\infty^B(n)+\gamma S_B(n)\ ,
\end{eqnarray}
and using Eqs. (\ref{fa}) and (\ref{perc}) we obtain the new fraction
of nodes $\mu_B(n)$. The process continues until the system reaches
the steady state at which $P_\infty^{A/B} < \epsilon$, where
$\epsilon=1/N$ is related to finite size effects. We iterate this
theoretical process numerically.

Figure \ref{figura_1} compares the simulation results (symbols) with
the theoretical results (dashed line). Note that both agree. Only when
$\gamma$ values are close to 1 and near $p_c$ do the simulation
results differ slightly from the theoretical results. To explain this
we compute theoretically the number of iterations (NOI) or stages
required for the system to reach the steady state (see
Fig. \ref{figura_3}). We can see that the NOI generates a sharp peak
at the critical threshold. The system requires many time steps to
reach the steady state when $p$ is close to $p_c$. When $p$ moves away
from $p_c$, only a few time steps are needed for the system to reach
the steady state. The internal structures of the two new GCs in B
(from simulation and theory) differ after we apply the strategy. After
a few steps they remain similar at the steady state for values of $p$
far from $p_c$. With internal structure we are referencing to the
number and form in which are made the internal connections of each
GC. Note that theoretically we obtain a new GC and the internal
connections change, but in the simulations we only add two new
connections to each saved finite cluster. For values of $p$ close to
$p_c$ the NOI increases exponentially and the difference between both
GCs increases. Finally the deviation becomes less noticeable as we
move away from $\gamma=1$ and near $p_c$ where the NOI is high. This
is the case because the probability of saving any finite cluster
decreases as $\gamma$ decreases, which means that only a few nodes are
saved and the restructuring of the GC is slight.

\begin{figure} 
\vspace{1cm}
\begin{center}
  \includegraphics[width=0.48\textwidth]{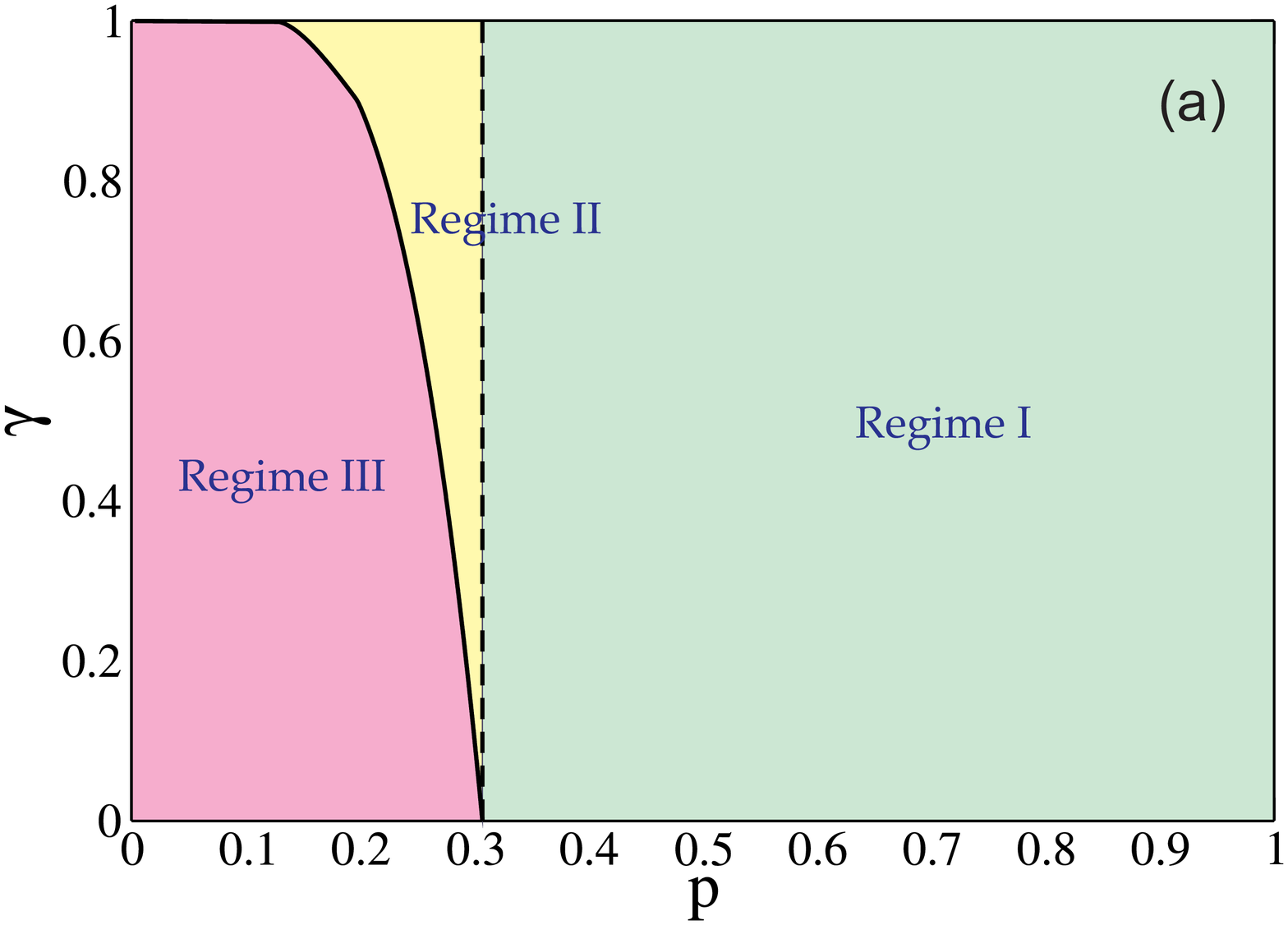}
  \includegraphics[width=0.48\textwidth]{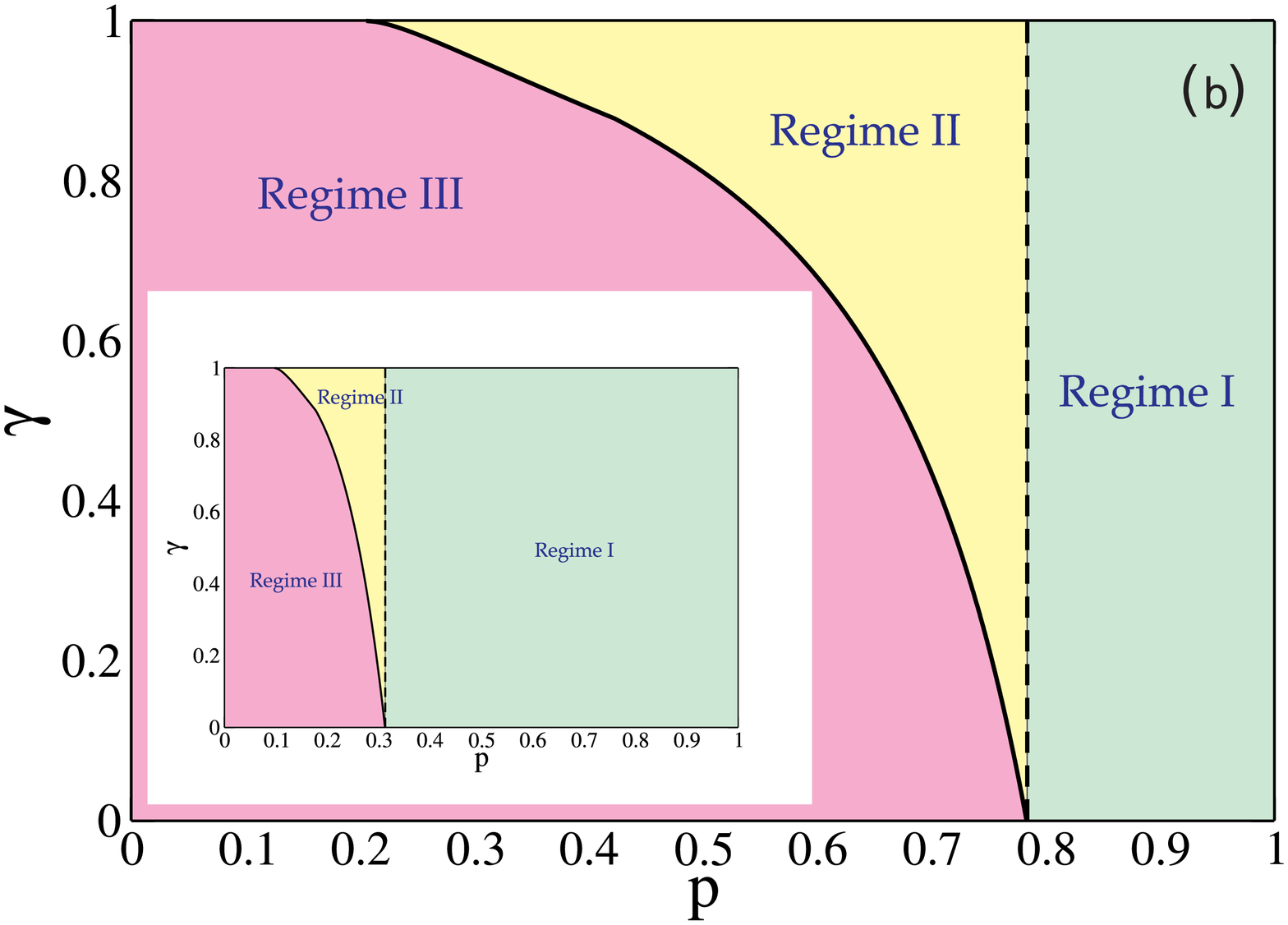}\\
  \includegraphics[width=0.48\textwidth]{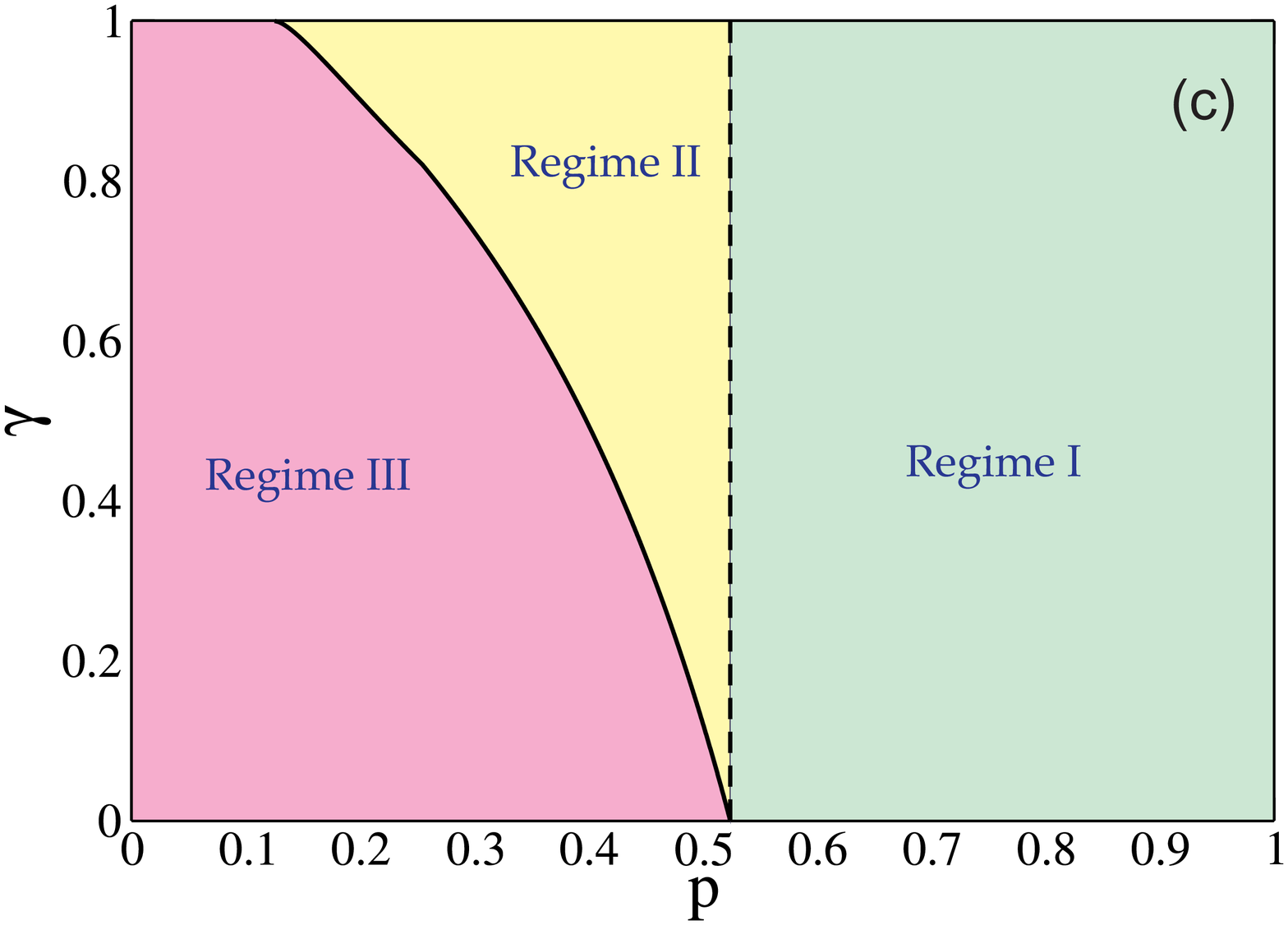}
  \includegraphics[width=0.48\textwidth]{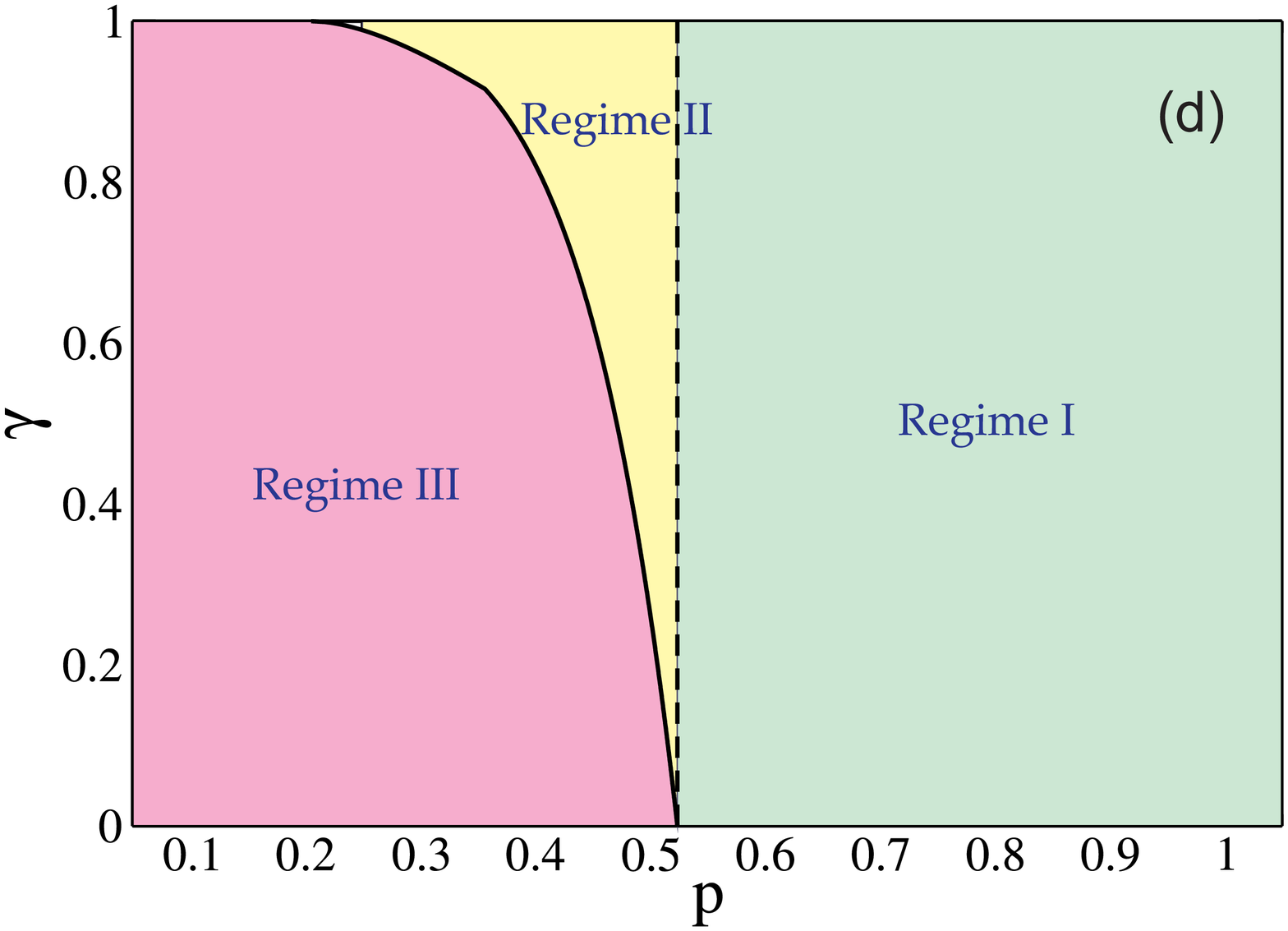}
\end{center}
\caption{Phase diagram in the plane $\gamma-p$ from the theory for
  $a)$ two ER with $\langle k\rangle=8$, $b)$ two SF with $\langle
  k\rangle=3.08$ and the combination of and ER and SF networks
  applying the strategy in $c)$ SF and $d)$ ER. The continuous curve
  represents the value of $\gamma_c$ below which the system is
  completely destroyed and the dashed line represents $p_c$ for the
  case $\gamma=0$. In the inset of $b)$ we plot the case for two SF,
  but with $kmin=5$ ($\langle k\rangle=8$). Notice that the plot is
  very similar to the two ER case.}\label{figura_4}
\end{figure}

Figure \ref{figura_4} plots the phase diagram in the plane
$\gamma$-$p$ for the same cases as in Fig.~\ref{figura_1}: (a) two ER
with $\langle k\rangle=8$, (b) two SF with $\langle k\rangle=3.08$,
and combinations of (c) ER-SF and (d) SF-ER where the strategy is
always applied to network B (A -- B). The vertical line (dashed) is
the $p_c$ for $\gamma=0$. To the right of the dashed line in Regime I
the system remains functional, even when the strategy is not
applied. The middle region, Regime II, is the zone in which the
strategy is needed to avoid the total destruction of the system. Note
that this region is much larger when the networks have a low average
degree, e.g., the two SF with $\langle k\rangle=3.08$. Figure
\ref{figura_4}(b) (inset) plots the phase diagram for two SF with
$\langle k\rangle=8$ ($k_{\rm min}=5$). Note that the plot is similar
to the one in Fig. \ref{figura_4}(a). Networks with a low average
degree have nodes with fewer connections, which means that they are
more prone to fragmentation and more likely to fail
completely. Figures \ref{figura_4}(c) and \ref{figura_4}(d) show that
in ER-SF and SF-ER Regime II is broader in ER-SF, and we are more able
to save the system when the strategy is applied to fragile networks,
i.e., those with a low average degree. This can be seen more clearly
in Fig. \ref{figura_3}(b) where for the same $\gamma$ the $p_c$ is
higher (blue curve) when the strategy is applied to the less fragile
network (ER). As mentioned above, the network with the lower average
degree is more prone to fragmentation, and its finite clusters
ultimately fail. It is thus important to apply the strategy to the
more fragile network with the lower average degree. Finally the
continuous curve represents the $\gamma_c$ values below which the
networks are completely destroyed (Regime III). We obtain this curve
theoretically from the peaks in the NOI (see Fig.~\ref{figura_3} for
different $\gamma$ values).

\section{Conclusions}

We have developed a strategy for avoiding the complete destruction of a
system of two interdependent networks. We define network
interconnections to be interdependent links connecting each node in the
first network with its counterpart in the second. We apply our strategy
to one network and prior to their failure and with probability $\gamma$
save every finite cluster by connecting two of their nodes to the GC. We
find that the system becomes increasingly robust to cascading failure as
$\gamma$ increases, and that the strategy is most effective when it is
applied to the network with the lower average degree. We solve the
problem theoretically using percolation theory and we find an agreement
with the simulation results.

In future work we will study the implementation of the strategy in
systems larger than two networks and where network interdependence is
given by a degree distribution.

\acknowledgments

We acknowledge UNMdP, FONCyT (Pict 0429/2013 and Pict 1407/2014) and
CONICET (PIP 00443/2014) for financial support. CELR acknowledges
CONICET for financial support. Work at Boston University is supported
by NSF Grants PHY-1505000, CMMI-1125290, and CHE-1213217, and by DTRA
Grant HDTRA1-14-1-0017.


\end{document}